\documentclass[aps, reprint,pra,longbibliography]{revtex4-1}
\usepackage{dcolumn}
\usepackage[utf8]{inputenc}
\usepackage{amsmath}
\usepackage{amsfonts}
\usepackage{amssymb}
\usepackage{graphicx}
\usepackage{hyperref}
\usepackage{xcolor}
\usepackage{lipsum}

\newcommand{\ket}[1]{\ensuremath{\left|{#1}\right\rangle}}
\newcommand{\bra}[1]{\ensuremath{\left\langle{#1}\right |}}

\begin{document}

\title{Relative Phase Distribution and the Precision of Optical Phase Sensing in Quantum Metrology}

\author{Felipe F. Braz}
\affiliation{Departamento de F\'isica, Universidade Federal de Minas Gerais, Belo Horizonte, MG 31270-901, Brazil}
\author{Tam\'iris R. Calixto}
\affiliation{Departamento de F\'isica, Universidade Federal de Minas Gerais, Belo Horizonte, MG 31270-901, Brazil}
\author{Pablo L. Saldanha}\email{saldanha@fisica.ufmg.br}
\affiliation{Departamento de F\'isica, Universidade Federal de Minas Gerais, Belo Horizonte, MG 31270-901, Brazil}

\date{\today}

\begin{abstract}
One of the quantum metrology goals is to improve the precision in the measurement of a small optical phase introduced in one optical mode in an interferometer, \textit{i.e.}, phase sensing. In this paper, we obtain the relative phase distribution introduced by Luis and Sánchez-Soto (LSS) [Phys. Rev. A \textbf{53}, 495 (1996)] for several two-mode pure quantum light states useful in quantum metrology. We show that, within the numerical precision of our calculations, the Fisher information obtained from the LSS relative phase distribution is equal to the quantum Fisher information for the considered states (the average difference for the tested states is smaller than 0.1\%). Our results indicate that the LSS relative phase distribution can be used to predict the minimum uncertainty possible in the process of phase sensing in quantum metrology, since this uncertainty depends on the quantum Fisher information, at least for pure states.
\end{abstract}


\maketitle

\section{Introduction}

The corpuscular nature of light imposes limits for the maximum possible precision in the estimation of a phase difference introduced between two optical modes in an interferometer. This quantum behavior generates fluctuations in the photon counts by detectors at the interferometer  exits, the so-called shot noise, which disturbs the phase estimation. With classical light having an average number of photons $\bar{N}$, the uncertainty in the phase sensing scales with $1/\sqrt{\bar{N}}$. But with quantum light sources, involving entanglement or light squeezing, this uncertainty may scale with $1/\bar{N}$ in some cases, such that a much greater precision can be achieved with a large value for $\bar{N}$ \cite{giovannetti04,dobr15,polino20}. 

The field of photonic quantum metrology has been dealing with this issue, and many recent experimental advances were reached \cite{dobr15,pirandola18,polino20}. The combination of entanglement, multiple samplings of the phase shift, and adaptive measurement have been use to optimize a phase shift estimation \cite{daryanoosh18}. Coherent measurements involving entangling operations can be useful for phase estimation in the presence of noise \cite{piera21}. Notable advances in multiple phases estimation have been reported \cite{guo20,gessner20,valeri20,hong21,liu21}. The use of detection schemes with photon-number-resolution allows the realization of quantum sensing protocols without pos-selection, such as scalable protocols for quantum-enhanced optical phase estimation \cite{you21} and distributed quantum sensing \cite{zhao21}. Recently it was also shown how the insensitivity of Hong-Ou-Mandel two-photon interference \cite{hong87} to phase fluctuations can be used to reduce the phase noise in the measurement of a mirror tilting angle \cite{aguilar20} in relation to a similar classical procedure \cite{walborn20}, increasing the precision. Quantum metrology with cavities and resonators are also interesting possibilities \cite{liu16,tang22}. But perhaps the most prominent application of photonic quantum metrology so far has been the improvement of the sensitivity of gravitational wave detectors \cite{ligo11,aasi13,zhao20,Sequino21}. It is worth mentioning that many other physical systems, besides quantum light, are used in the broader area of quantum metrology \cite{ma11,pezze18,xu22}.


Here we associate the precision limit in the estimation of a small phase difference introduced between two optical modes to the phase difference distribution of the initial two-mode  quantum light state used for this purpose. For many pure quantum states useful in quantum metrology, we compute the relative phase distribution $P(\phi)$ introduced by Luis and Sánchez-Soto (LSS) \cite{luis96}. We then compute the Fisher information based on this probability distribution and show that the result is very close to the quantum Fisher information for the treated states. The average difference between these quantities for the tested quantum states is smaller than 0.1\%, a difference compatible with the numerical precision of our calculations. Since the quantum Fisher information is associated to the maximum possible precision in the phase estimation according to the rules of quantum mechanics \cite{braunstein94,dobr15,polino20,pezze18,sidhu20}, our results demonstrate the relevance of the LSS relative phase distribution in the field of quantum metrology. If the introduced method can be extended to treat mixed states, the LSS relative phase distribution could be a valuable tool to estimate the maximum precision in phase sensing for realistic situations involving mixed states subjected to decoherence processes, since this is usually a difficult task with the use of the quantum Fisher information \cite{escher11}.

We consider in this work NOON states \cite{bollinger96}, phase states \cite{sanders95}, states produced with the incidence of a twin-Fock state \cite{holland93} and with a correlated Fock state \cite{yurke98,dowling98} at the interferometer inputs, and with the incidence of a squeezed state in one interferometer input and a coherent state in the other \cite{caves81}, besides the ``classical'' situation of a Fock state sent at one of the interferometer inputs. Our results give some insights for the fundamental reason behind the improved sensitivity of the phase estimation by using quantum light sources.


\section{Precision limit in phase sensing}

The situation we want to discuss in this paper is depicted in Fig. 1(a). A general two-mode optical pure state is prepared by a source, which can be written as 
\begin{equation}\label{psi0}
	\ket{\Psi}=\sum_{N,k} A_{N,k} \ket{k}_a\ket{N-k}_b,
\end{equation}
with $\ket{k}_a$ representing a Fock state with $k$ photons in mode $a$ and $\ket{N-k}_b$ a state with $N-k$ photons in mode $b$ ($N$ is the total number of photons). During the system evolution, a small relative phase $\theta$ is introduced between modes $a$ and $b$, such that the state after the evolution is $\ket{\Psi_\theta}=e^{i \hat{n}_a\theta}\ket{\Psi}$, where $\hat{n}_a$ is the photon-number operator for mode $a$. Measurements on the final state are performed with the objective to estimate $\theta$, whose value is initially unknown. The procedure can be repeated $p$ times to improve the precision. 

The limit in precision for a given measurement strategy is given by the Cram\'er-Rao bound \cite{dobr15,polino20,pezze18,sidhu20}
\begin{equation}\label{CRB}
	\Delta \theta \ge  \frac{1}{\sqrt{p F(\theta)}},
\end{equation}
where
\begin{equation}\label{Fish-gen}
	F(\theta)=\int dx \frac{1}{P(x,\theta)}\left[\frac{dP(x,\theta)}{d\theta}\right]^2
\end{equation}
is the Fisher information on $\theta$ for a given measurement configuration, $P(x,\theta)$ being the probability distribution of obtaining a measurement result $x$ when the parameter to be estimated is $\theta$. We consider a continuous distribution for the possible experimental results, but the case of discrete values is completely analogous. The maximization of $F(\theta)$ over all possible quantum measurements yields the quantum Fisher information $F_Q$ for the quantum state $\ket{\Psi}$ under the evolution $e^{i \hat{n}_a\theta}$, which leads to the ultimate precision bound  \cite{braunstein94,dobr15,polino20,pezze18,sidhu20}. The quantum Fisher information for this kind of evolution with pure states is given by \cite{dobr15,polino20}
\begin{equation}\label{Fisher}
	F_Q=4\Delta^2 n_a=4\left[ \bra{\Psi} \hat{n}_a^2\ket{\Psi} - (\bra{\Psi} \hat{n}_a\ket{\Psi})^2 \right].
\end{equation}
According to the rules of quantum mechanics, the minimum possible uncertainty in the estimation of the phase $\theta$ using a quantum state $\ket{\Psi}$ under the evolution $e^{i \hat{n}_a\theta}$ is then
\begin{equation}\label{CRBmin}
	\Delta \theta_\mathrm{min} =  \frac{1}{\sqrt{p F_Q}},
\end{equation}
with $F_Q$ given by Eq. (\ref{Fisher}).


\begin{figure}
  \centering
    \includegraphics[width=8.5cm]{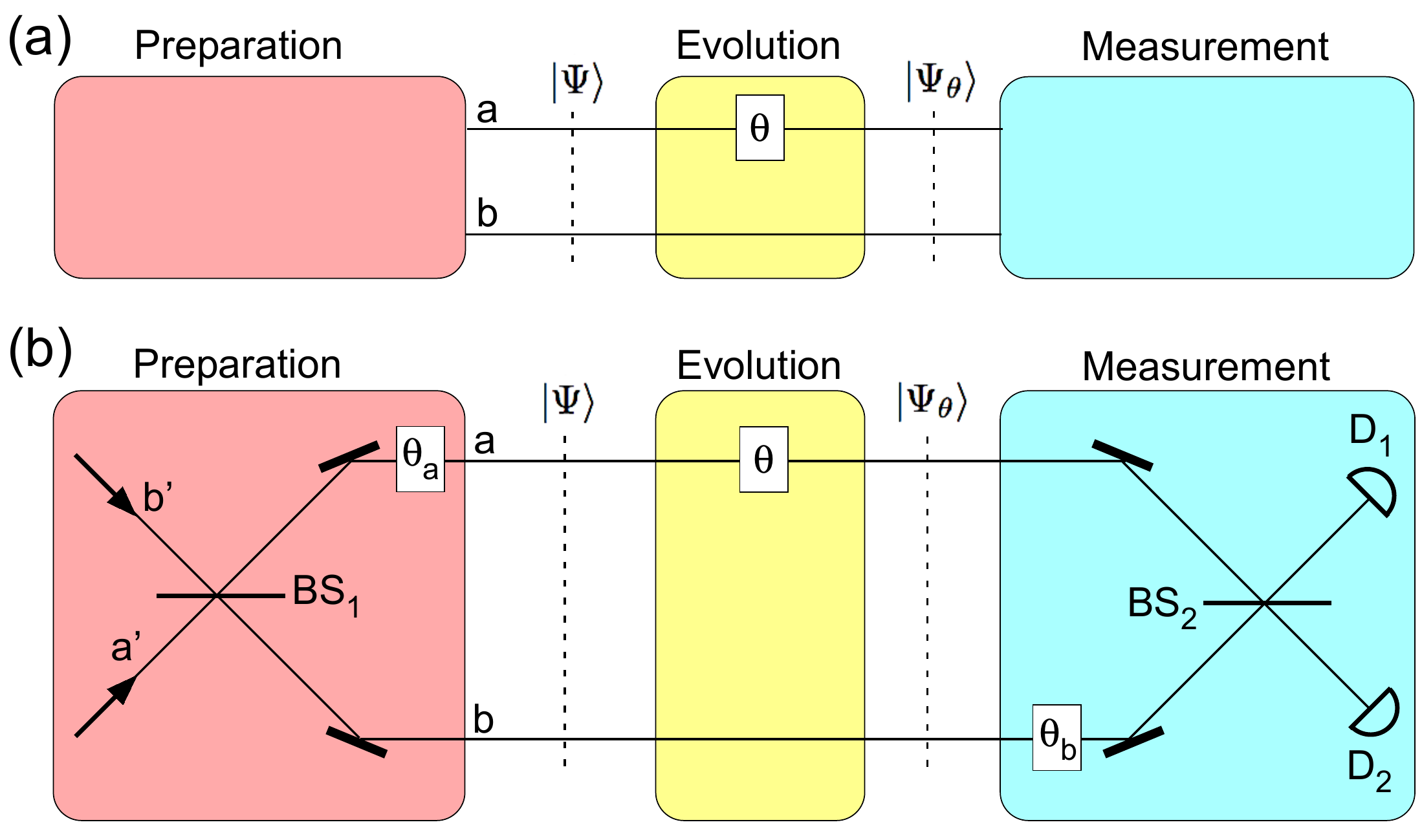}
  \caption{(a) General scheme for the estimation of a small phase $\theta$ introduced in one optical mode. An initial two-mode state $\ket{\Psi}$ is prepared. The state evolves to $\ket{\Psi_\theta}$ due to the introduced phase. Measurements are performed in the final state to estimate $\theta$. (b) A possible way to implement the protocol. Light in a known separable state in modes $a'$ and $b'$ is sent to the input ports of the beam splitter BS$_1$. After this beam splitter, a phase $\theta_a$ is added in mode $a$, preparing the state $\ket{\Psi}$ in modes $a$ and $b$. After the system evolution, a phase $\theta_b$ is introduced in mode $b$ before the paths recombine in beam splitter BS$_2$. Detectors D$_1$ and D$_2$ measure the intensity (or the number of photons) at the interferometer exits and $\theta$ can be estimated from these results.}
\end{figure}

Fig. 1(b) illustrates a possible way to prepare the initial state and to measure the final state, with the system composing a Mach-Zehnder interferometer. Light in a known two-mode separable quantum state is sent to the input ports of the beam splitter BS$_1$, and right after this beam splitter a phase $\theta_a$ is added to prepare the state $\ket{\Psi}$. Later we will use this phase $\theta_a$ to let the mean of the relative phase distribution in zero. A known phase $\theta_b$ can be included in path $b$ before the paths to be combined in beam splitter BS$_2$. $\theta_b$ can be chosen in order that the inclusion of  a small phase $\theta$ produces a maximum variation in the statistics of the number of photons detected by the photon detectors, such that this phase can be properly estimated. In some situations to be treated in this work we consider the quantum states before BS$_1$ as in Fig. 1(b), but in others we start from the state $\ket{\Psi}$ as in Fig. 1(a).

With the use of ``classical'' light states (such as coherent states) with an average number of photons $\bar{N}$ in an interferometer, the quantum Fisher information is, in the best scenario, $F_Q=\bar{N}$. From Eq. (\ref{CRBmin}), we have $\Delta \theta_\mathrm{min} = 1/\sqrt{p\bar{N}}$, with a phase estimation uncertainty limited by the shot-noise limit. But for some quantum states with an average number of total photons $\bar{N}$, we have $F_Q=\bar{N}^2$ in Eq. (\ref{Fisher}), such that from Eq. (\ref{CRBmin}) we have $\Delta \theta_\mathrm{min}= 1/(\bar{N}\sqrt{p})$, the so-called Heisenberg limit, the maximum precision permitted by quantum mechanics \cite{dobr15,polino20,pezze18,sidhu20}.


\section{Relative phase distribution}

There is a uncertainty relation between the phase and the number of photons in an optical mode. But since there is no phase operator in quantum optics, this is not an uncertainty that comes from the non-commutative behavior of operators. It is similar to the uncertainty relation between energy and time, as we also do not have a time operator in quantum mechanics. Quantum optical states with a relatively well defined phase must have a large uncertainty in the number of photons and states with a well defined number of photons have a completely undetermined phase \cite{pegg88,mandel,gerry}. 

One of the difficulties for establishing a quantum operator for the optical phase is that the spectrum of eigenvalues of the ``conjugate quantity'', the number of photons, does not have negative values. But usually one is not interested in an absolute phase, but in the relative phase between two optical modes. The ``conjugate quantity'' in this case is the difference between the number of photons in each of these modes, which can assume negative values. LSS defined an operator $\hat{E}$ with eigenvectors $\ket{\phi^{(N)}}$ in this two-mode space with the properties
\begin{equation}
	\hat{E}\ket{\phi^{(N)}}=e^{i\phi}\ket{\phi^{(N)}},\;\;\hat{n}\ket{\phi^{(N)}}=N\ket{\phi^{(N)}},
\end{equation}
where $\hat{n}=\hat{n}_a+\hat{n}_b$ is the operator for the total number of photons in the modes $a$ and $b$ \cite{luis93}. On this way $\hat{E}$ is equivalent to the ``imaginary exponential of the phase difference''. $\hat{E}$ is not Hermitian and commutes with $\hat{n}$, such that the states $\ket{\phi^{(N)}}$ are eigenstates of these two operators. These eigenvectors can be written as
\begin{equation}\label{phiN}
	\ket{\phi^{(N)}}=\frac{1}{\sqrt{2\pi}}\sum_{k=0}^Ne^{ik\phi} \ket{k}_a\ket{N-k}_b.
\end{equation}
The phase states $\ket{\phi^{(N)}}$ appear with other definitions for the phase-difference operator \cite{barnett90} and are also associated to optimum phase measurements in interferometers \cite{sanders95,luis96b}. 

In a later work, LSS defined the following probability distribution for the relative phase between two optical modes in the quantum state $\rho$ \cite{luis96}:
\begin{equation}\label{Prho}
	P(\phi)=\sum_{N=0}^{\infty}\bra{\phi^{(N)}}\rho\ket{\phi^{(N)}},
\end{equation}
with $\ket{\phi^{(N)}}$ given by Eq. (\ref{phiN}). Here we consider pure quantum states with a fixed number of total photons $N$:
\begin{equation}\label{psin}
	\ket{\Psi_N}=\sum_{k} A_{N,k} \ket{k}_a\ket{N-k}_b.
\end{equation} For these states, the probability distribution reduces to 
\begin{equation}\label{PN}
	P_N(\phi)=|\langle\phi^{(N)}|\Psi_N\rangle|^2,
\end{equation}
with $\ket{\phi^{(N)}}$ given by Eq. (\ref{phiN}). This probability distribution gives an idea of how well defined is the relative phase between the modes for this quantum state. We have a  uncertainty relation between the phase difference and the difference in the number of photons between the two modes. If the photon number difference is well known, we have a uniform distribution for the phase difference. For distributions with a well localized phase difference, the photon number difference has a large uncertainty. Calling $\Delta\phi$ the width of the distribution of the phase difference and $\Delta n_-$ the width in the distribution of the photon difference, we have $\Delta\phi\Delta n_-\geq \pi$ in general (considering $\Delta n_-\geq1/2$). For states with a fixed total number of photons $N$, we can write $n_-=n_a-n_b=2n_a-N$, such that $\Delta n_-=2\Delta n_a$ and we have
\begin{equation}
	\Delta\phi\Delta n_a\geq \frac{\pi}{2}.
\end{equation}

We may compare the above inequality with the one obtained from Eq. (\ref{CRB}) for $p=1$ substituting $F(\theta)$ by $F_Q$ and using Eq. (\ref{Fisher}):
\begin{equation}
	\Delta\theta\Delta n_a\geq \frac{\pi}{2}.
\end{equation} 
Note that the uncertainties $\Delta\phi$ and $\Delta\theta$ obey the same inequality with $\Delta n_a$. But in principle they represent different quantities: $\Delta\phi$ is the phase difference uncertainty of a quantum state $\ket{\Psi_N}$ according to the distribution defined in Eq. (\ref{PN}), while $\Delta\theta$ is the precision in the estimation of a small optical phase $\theta$ in the scheme of Fig. 1(a) when the initial state is $\ket{\Psi_N}$. But it is clear that these two quantities must be related. To be possible to estimate a deviation $\delta\theta$ in the scheme of Fig. 1(a), this deviation must displace the initial phase distribution of the state $\ket{\Psi_N}$ by an amount greater than its initial uncertainty, \textit{i.e.}, we must have $\delta\theta\geq \Delta\phi$. So the minimum detectable value of $\delta\theta$ (which is roughly the uncertainty $\Delta\theta$ in the phase estimation) should be approximately  $\Delta\phi$,\textit{ i.e.}, $\Delta\theta\approx\Delta\phi$. But note that, for periodic phase difference distributions, a phase deviation smaller than $\Delta\phi$, of the order of the distribution period, may change it much. So, a more robust way to define the perturbation generated by a small phase added to the relative phase distribution is necessary.  

Let us consider the classical fidelity (often called Bhattacharyya fidelity) between two nearby probability distributions \cite{pezze18}:
\begin{equation}
	\mathcal{F}=\int d\phi \sqrt{P_N(\phi)P_N(\phi+\delta\phi)},
\end{equation}
where $P_N(\phi)$ is the relative phase distribution of Eq. (\ref{PN}). By considering terms up to $(\delta\phi)^2$ i the Taylor expansion of the fidelity, we have
\begin{equation}
	\mathcal{F}\approx 1 - \frac{1}{8}F_\mathrm{LSS}(\delta\phi)^2,
\end{equation}
where
\begin{equation}\label{Fisher-LSS}
	F_\mathrm{LSS}=\int d\phi \frac{1}{P_N(\phi)}\left[\frac{dP_N(\phi)}{d\phi}\right]^2
\end{equation}
is the Fisher information for the LSS relative phase distribution $P_N(\phi)$ of Eq. (\ref{PN}). The Fisher information of Eq. (\ref{Fish-gen}) can be derived in the same way from the fidelity between the probability distributions of the measurement results for states differing by a small amount in the parameter $\theta$ \cite{pezze18,sidhu20}. In the next section we show that the difference between the Fisher information $F_\mathrm{LSS}$ above and the quantum Fisher information from Eq. (\ref{Fisher}) is typically smaller than 0.2\% for many different quantum states useful in quantum metrology, this small difference being compatible with the numerical precision of the calculations. 


\section{Relative phase distribution and the precision of phase sensing for useful states in quantum metrology}

We now proceed to show the equivalence, under the numerical precision of our calculations, between the Fisher information $F_\mathrm{LSS}$ from Eq. (\ref{Fisher-LSS}), based on the LSS relative phase distribution of Eq. (\ref{PN}), and the quantum Fisher information $F_Q$ from Eq. (\ref{Fisher}). We consider many families of quantum states useful in quantum metrology.

\subsection{Fock state at one interferometer input}

The first class of states we treat is the one with a Fock state $\ket{N}_{a'}$ in one of the interferometer inputs of Fig. 1(b) and a vacuum state in the other. This is a case with no advantage in relation to the use of a ``classical'' coherent state. BS$_1$ performs the following transformation on the annihilation operators for the considered modes:
\begin{equation}\label{BS}
	\hat{a}'\rightarrow \frac{1}{\sqrt{2}}(\hat{a}-i\hat{b}),\;\;\hat{b}'\rightarrow \frac{1}{\sqrt{2}}(-i\hat{a}+\hat{b}).
\end{equation}
Considering these transformations and an introduced phase $\theta_a=-\pi/2$ in the scheme of Fig. 1(b), the quantum state $\ket{\Psi_N}$ inside the interferometer is given by Eq. (\ref{psin}) with coefficients
\begin{equation}\label{fock}
	A_{N,k}=\frac{i^{N-k}}{\sqrt{2^{N}}}\binom{N}{k}\bigg/\sqrt{\binom{N}{k}},
\end{equation}
whose squared modulus are shown in Fig. 2(a) for $N=10$. The relative phase distribution of Eq. (\ref{PN}) is shown in Fig. 2(b) for $N=10$. In this situation, $F_Q$ from Eq. (\ref{Fisher}) and $F_\mathrm{LSS}$ from Eq. (\ref{Fisher-LSS}) can be computed analytically, both being equal to $N$. Fig. 2(c) shows the analytical values for $F_Q$ (continuous gray line) and the numerically computed values for $F_\mathrm{LSS}$ (black circles), as a function of $N$. The average percentage difference between these quantities for the points shown in Fig. 2(c) is 0.00001\%.

\begin{figure}
    \centering
    \includegraphics[width=8.5cm]{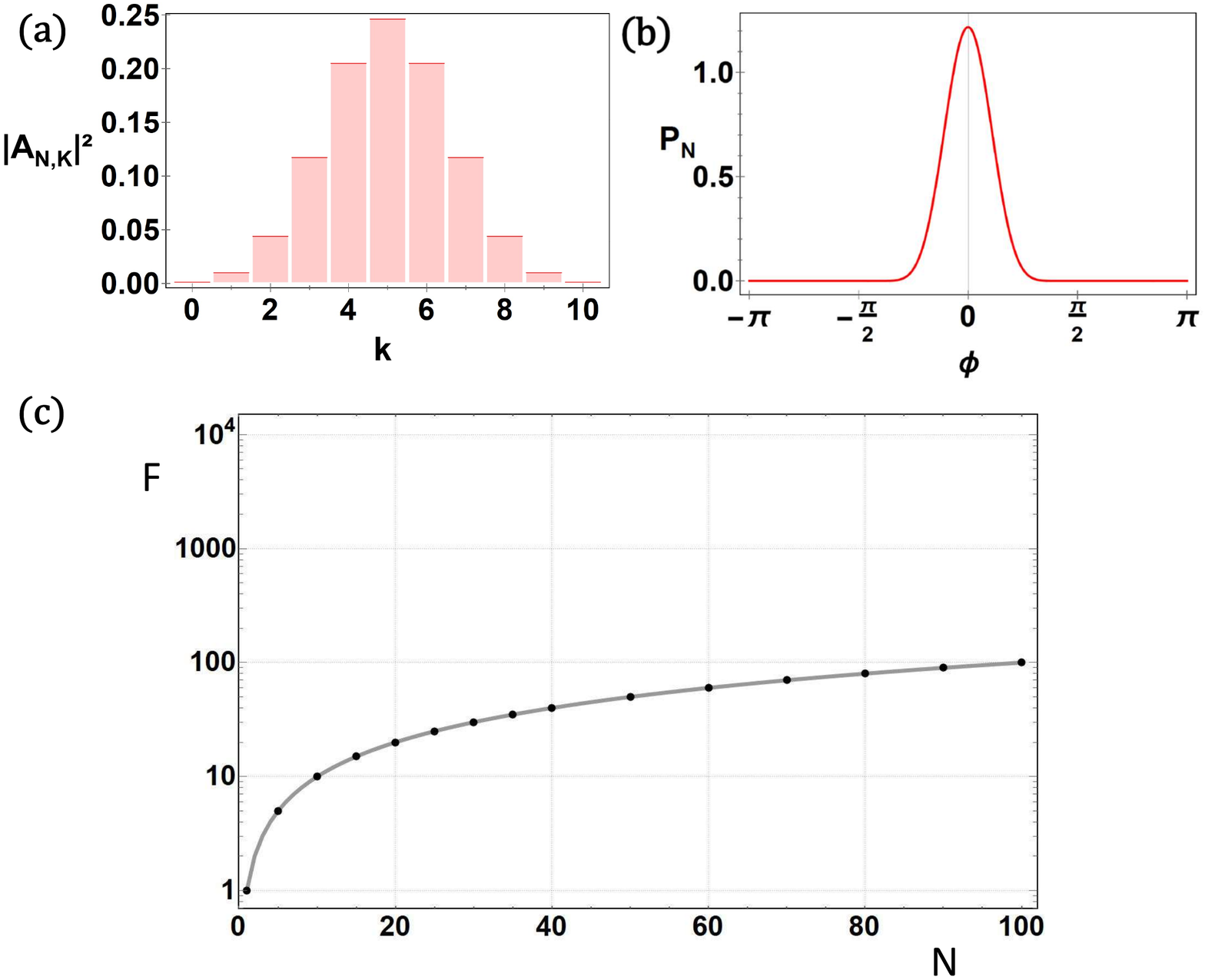}
    \caption{Analysis of Fock states at one of the inputs of the interferometer of Fig. 1(b), resulting in the state of Eq. (\ref{psin}) with the coefficients of Eq. (\ref{fock}). (a) Squared modulus of the coefficients from Eq. (\ref{fock}) with $N=10$. (b) Relative phase distribution from Eq. (\ref{PN}) for $N=10$. (c) $F_Q$ from Eq. (\ref{Fisher}) (continuous gray line), which in this case is $F_Q=N$, and $F_\mathrm{LSS}$ from Eq. (\ref{Fisher-LSS}) (black circles), which was numerically computed, as a function of $N$.}
    \label{fig:fock-vacuo}
\end{figure}

\subsection{NOON state}

The second class of states we treat is the one with a NOON state \cite{bollinger96} in modes $a$ and $b$ in Fig. 1(a):
\begin{equation}\label{NOON}
	\ket{\Psi_N}=\frac{1}{\sqrt{2}}\left[ \ket{N}_a\ket{0}_b + \ket{0}_a\ket{N}_b  \right].
\end{equation}
The coefficients of Eq. (\ref{psin}) are $A_{N,k}=(\delta_{k,0}+\delta_{k,N})/\sqrt{2}$ in this case and their squared modulus are shown in Fig. 3(a) for $N=10$. The relative phase distribution of Eq. (\ref{PN}), which in this case is given by $P_N(\phi)=[1+\cos(N\phi)]/(2\pi)$, is shown in Fig. 3(b) for $N=10$. Note that the interference pattern of a NOON state has a periodicity $2\pi/N$, which is compatible with the $N$ peaks present in the relative phase distribution shown in Fig. 3(b). For a NOON state, $F_Q$ from Eq. (\ref{Fisher}) and $F_\mathrm{LSS}$ from Eq. (\ref{Fisher-LSS}) can be computed analytically, both being equal to $N^2$, resulting in the Heisenberg limit $\Delta\theta_\mathrm{min}=1/(N\sqrt{p})$ in Eq. (\ref{CRBmin}). Fig. 3(c) shows the behavior of $F_Q=N^2$ (continuous gray line) and the numerically computed value for $F_\mathrm{LSS}$ (black circles), as a function of $N$. The average percentage difference between $F_Q$ and $F_\mathrm{LSS}$ for the points shown in Fig. 3(c) is $0.14\%$. Since these quantities are equal for the NOON states, the differences are due to the numerical precision of our calculations. So, we conclude that an average difference of the order of $0.14\%$, due to the numerical precision of our calculations, would be expected for the other states we treat in the following.

\begin{figure}
  \centering
    \includegraphics[width=8.5cm]{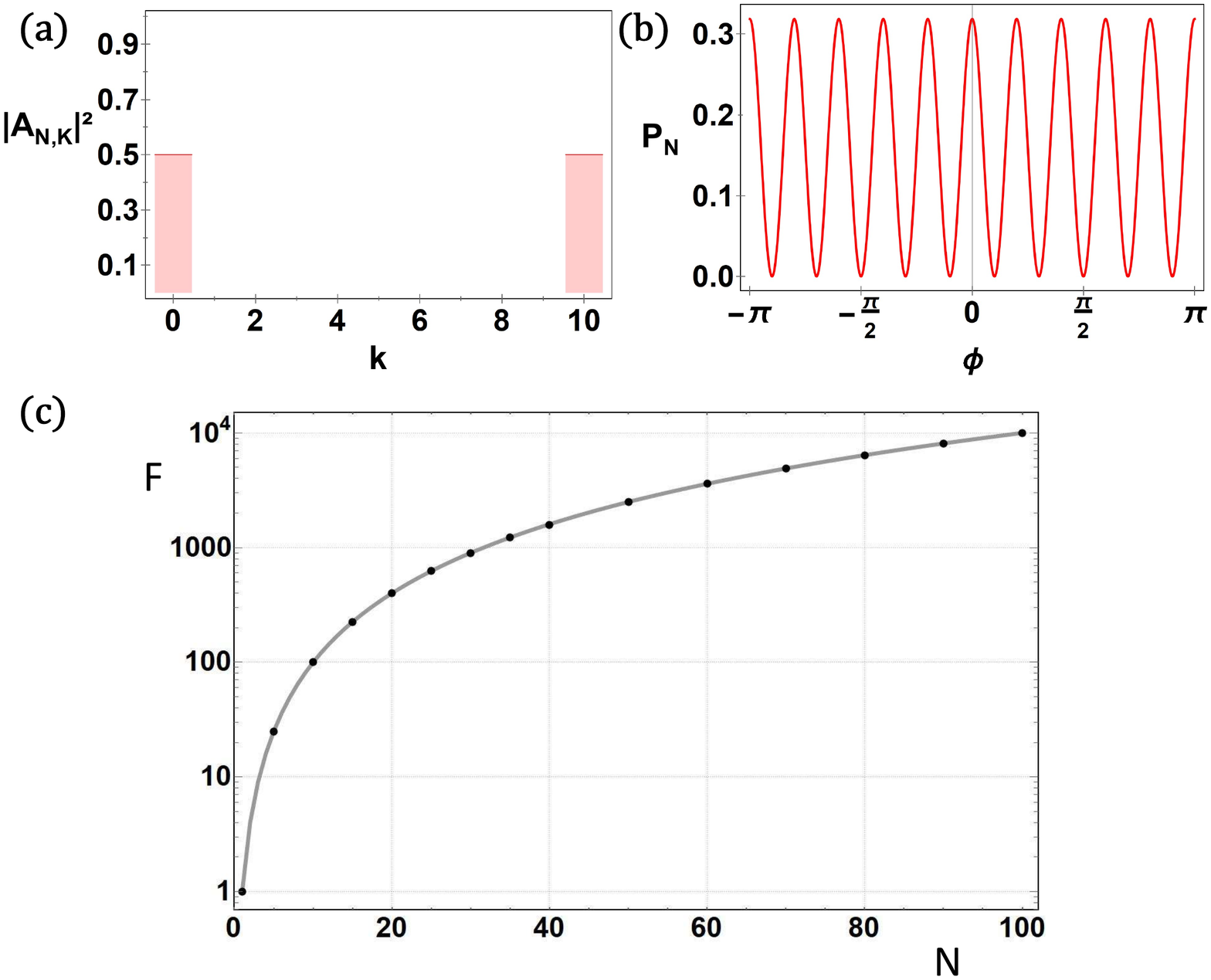}
  \caption{Analysis of the NOON states given by Eq. (\ref{NOON}). (a) Squared modulus of the coefficients defined in Eq. (\ref{psin}) for a NOON state with $N=10$. (b) Relative phase distribution from Eq. (\ref{PN}) for $N=10$. (c) $F_Q$ from Eq. (\ref{Fisher}) (continuous gray line), which in this case is given by $F_Q=N^2$, and $F_\mathrm{LSS}$ from Eq. (\ref{Fisher-LSS}) (black circles), which was numerically computed, as a function of $N$.}
\end{figure}

\subsection{Phase state}

The third class of states we treat is the one with a phase state $\ket{\phi^{(N)}}$ given by Eq. (\ref{phiN}) in modes $a$ and $b$ of Fig. 1(a) with $\phi=0$. The coefficients of Eq. (\ref{psin}) are $A_{N,k}=1/\sqrt{N+1}$ in this case and their squared modulus are shown in Fig. 4(a) for $N=10$. The relative phase distribution of Eq. (\ref{PN}), which in this case is given by $P_N(\phi)=\{\sin{[\phi(N+1)/2]}/\sin{[\phi/2]}\}^2/[2\pi(N+1)]$, is shown in Fig. 4(b) for $N=10$. Fig. 4(c) shows the behavior of $F_Q=(N^2+2N)/3$ (continuous gray line), which was analytically computed, and $F_\mathrm{LSS}$ (black circles), which was numerically computed, as a function of $N$. The average percentage difference between $F_Q$ and $F_\mathrm{LSS}$ for the points shown in Fig. 4(c) is $0.20\%$.

\begin{figure}
  \centering
    \includegraphics[width=8.5cm]{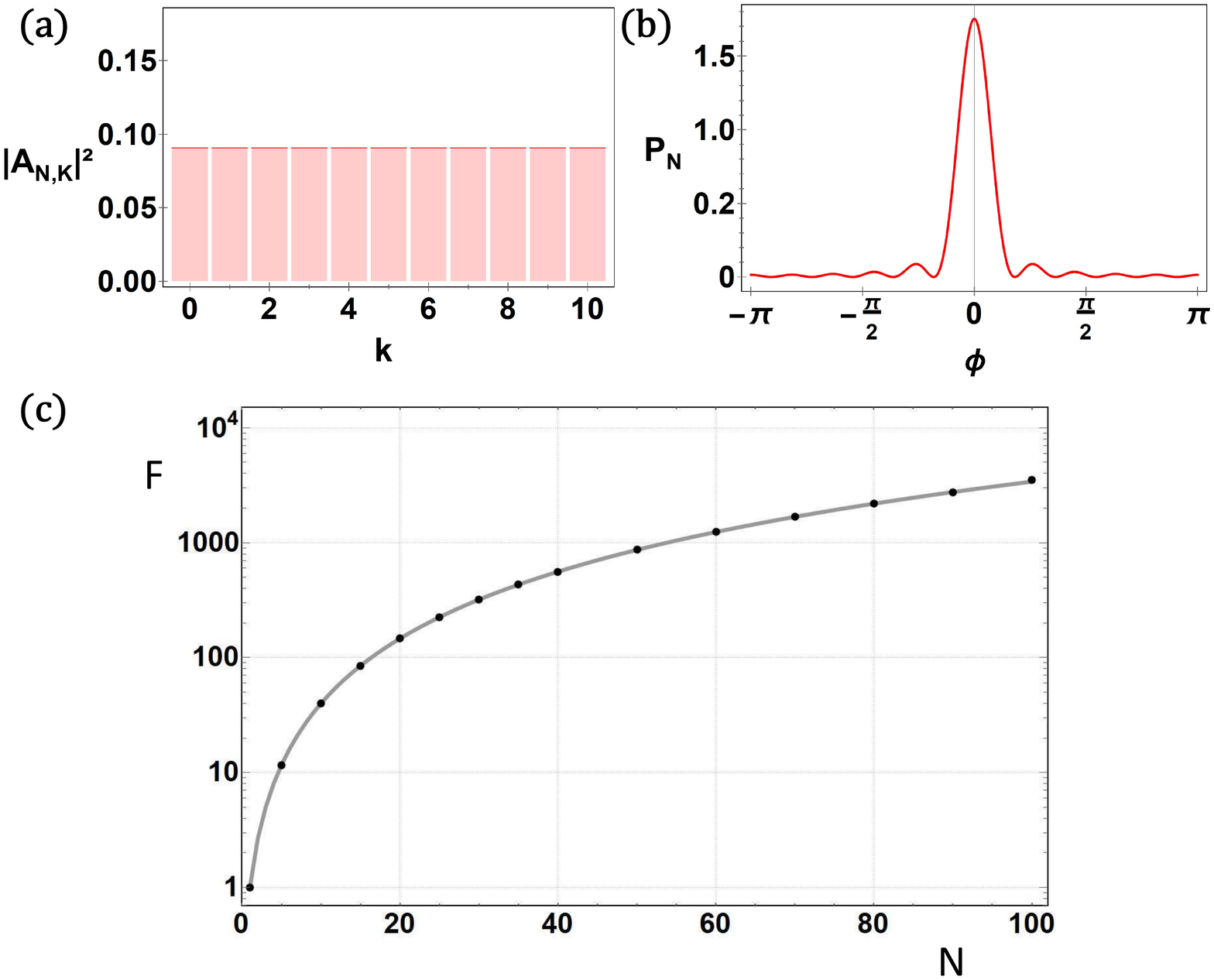}
  \caption{Analysis of the phase states given by Eq. (\ref{phiN}). (a) Squared modulus of the coefficients defined in Eq. (\ref{psin}) for a phase state with $N=10$. (b) Relative phase distribution from Eq. (\ref{PN}) for $N=10$. (c) $F_Q$ from Eq. (\ref{Fisher}) (continuous gray line), which in this case is given by $F_Q=(N^2+2N)/3$, and $F_\mathrm{LSS}$ from Eq. (\ref{Fisher-LSS}) (black circles), which was numerically computed, as a function of $N$.}
\end{figure}

\subsection{Twin-Fock state at the interferometer inputs} \label{sec-twin-Fock}

The next class of states we treat is the one where at the input of the interferometer of Fig. 1(b) we have the state $\ket{N/2}_{a'}\otimes\ket{N/2}_{b'}$, with two Fock states with $N/2$ photons in each mode \cite{holland93} ($N$ is even, naturally). In this case, the phase $\theta_a$ in Fig. 1(b) is set to $\theta_a=0$ in order to place the central peak of the relative phase distribution in $\phi=0$. The coefficients of Eq. \eqref{psin} for the state $\ket{\Psi_N}$ in modes $a$ and $b$ after BS$_1$ are then
\begin{equation}\label{twin-fock}
\begin{split}
    A_{N,k}&=\frac{i^{N/2}}{\sqrt{2^N}}\sqrt{\binom{N}{N/2}}\sum_{q=0}^{N/2}i^{-2q+k}\times\\&\times\binom{N/2}{q}\binom{N/2}{k-q}\bigg/\sqrt{\binom{N}{k}}
\end{split}
\end{equation}
and their squared modulus are shown in Fig. 5(a) for $N=10$. The relative phase distribution of Eq. (\ref{PN}) is shown in Fig. 5(b) for $N=10$. Fig. 5(c) shows the behavior of $F_Q$ (continuous gray line) and $F_\mathrm{LSS}$ (black circles), which were numerically computed, as a function of $N$.  The red traced line illustrates the behavior shown in Ref. \cite{sahota15} for the quantum Fisher information of the twin-Fock state for large $N$, $N^2/2+N$. The average percentage difference between $F_Q$ and $F_\mathrm{LSS}$ for the points shown in Fig. 5(c) is $0.05\%$.

\begin{figure}
  \centering
    \includegraphics[width=8.5cm]{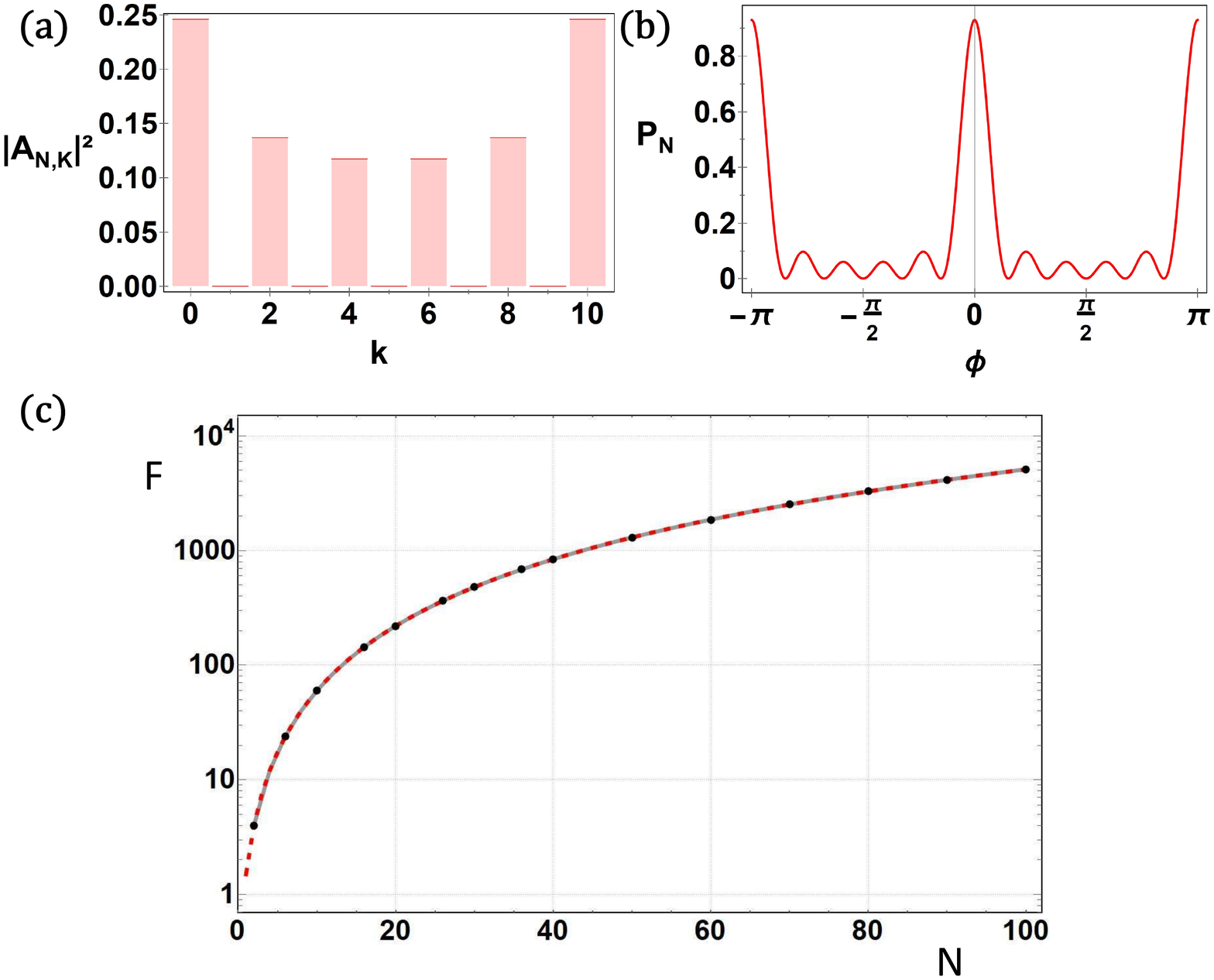}
  \caption{Analysis of the twin-Fock state at the inputs of the interferometer of Fig. 1(b), resulting in the state of Eq. (\ref{psin}) with the coefficients of Eq. (\ref{twin-fock}). (a) Squared modulus of the coefficients from Eq. (\ref{twin-fock}) with $N=10$. (b) Relative phase distribution from Eq. (\ref{PN}) for $N=10$. (c) $F_Q$ from Eq. (\ref{Fisher}) (continuous gray line) and $F_\mathrm{LSS}$ from Eq. (\ref{Fisher-LSS}) (black circles), both numerically computed, as a function of $N$. The red traced line shows the function  $N^2/2+N$.}\label{fig:fock-fock}
\end{figure}

\subsection{Correlated Fock state at the interferometer inputs}

\begin{figure}
  \centering
    \includegraphics[width=8.5cm]{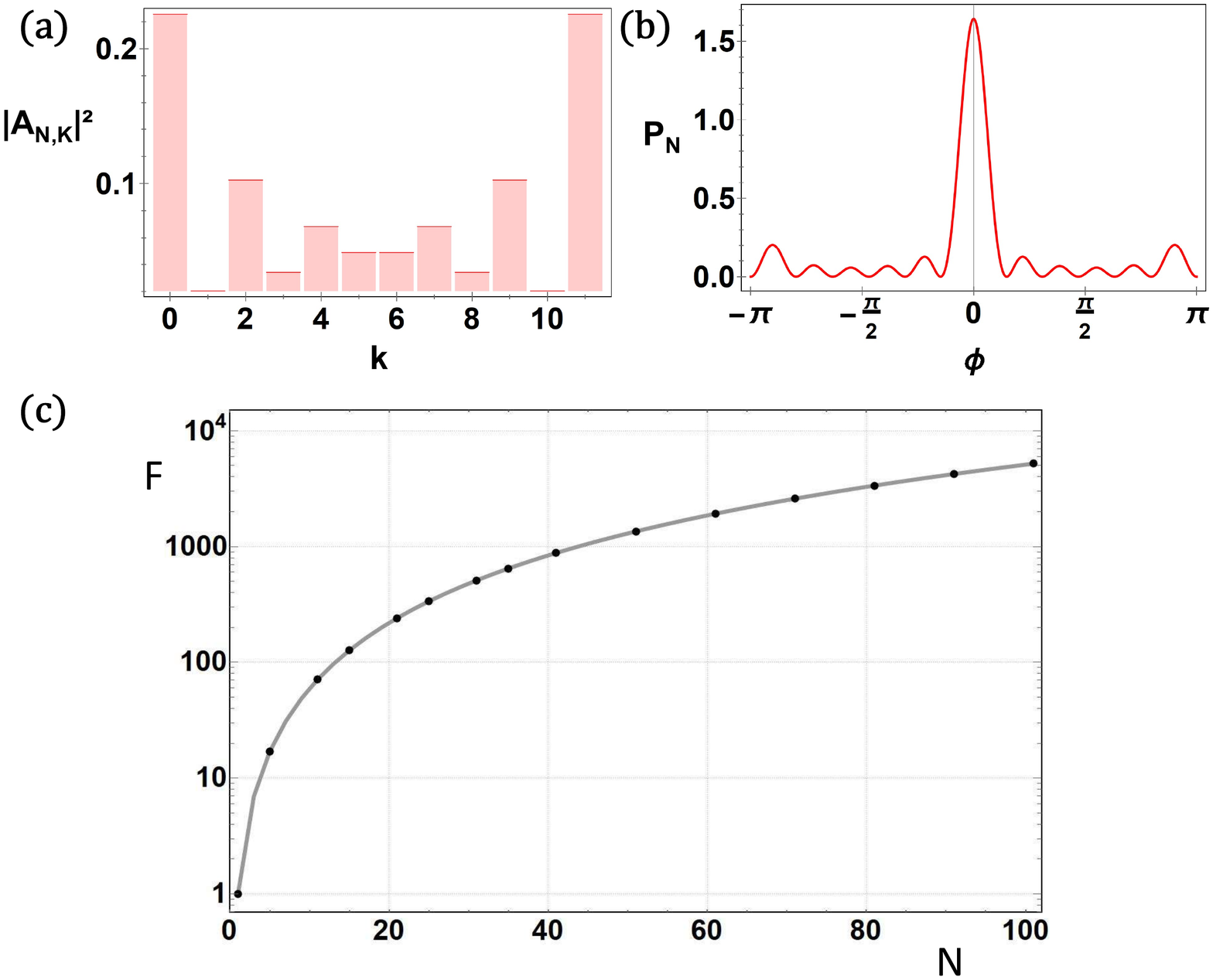}
  \caption{Analysis of the correlated Fock state at the inputs of the interferometer of Fig. 1(b), resulting in the state of Eq. (\ref{psin}) with the coefficients of Eq. \eqref{coef.entgl}. (a) Squared modulus of the coefficients from Eq. \eqref{coef.entgl} with $N=11$. (b) Relative phase distribution from Eq. (\ref{PN}) for $N=11$. (c) $F_Q$ from Eq. (\ref{Fisher}) (continuous gray line) and $F_\mathrm{LSS}$ from Eq. (\ref{Fisher-LSS}) (black circles), both numerically computed, as a function of $N$. }\label{fig:entgl}
\end{figure}

Another class of states we treat is with a correlated Fock state arriving at the input ports of the interferometer of Fig. 1(b) \cite{yurke98,dowling98}, of the form 
\begin{equation}\label{ent.state}
    \ket{\Psi_N}=\frac{1}{\sqrt{2}}\left[\ket{N_+}_{a'}\ket{N_-}_{b'}+\ket{N_-}_{a'}\ket{N_+}_{b'}\right],
\end{equation}
where $N_{\pm}=(N\pm1)/2$. In this case, $N$ must be odd. Again, the BS$_1$ will transform the annihilation operators according to Eq. \eqref{BS}. Here we also considered that the phase $\theta_a$ after the BS$_1$ in Fig. 1(b) is $\theta_a=0$. The coefficients for the quantum state inside the interferometer, written as in Eq. (\ref{psin}), are 
\begin{equation}\label{coef.entgl}
  \begin{split}
      A_{N,k}&=\frac{i^{N/2}}{\sqrt{2^{N-1}}}\sqrt{\binom{N}{N_{+}}}\sum_{q=0}^{N_+}\cos{\left[\frac{\pi}{4}(2k-4q+1)\right]}\times\\&\times\binom{N_+}{q}\binom{N-N+}{k-q}\bigg/\sqrt{\binom{N}{k}},
  \end{split}
\end{equation}
and their squared modulus are shown in Fig. \ref{fig:entgl}(a) for $N=11$. The relative phase distribution of Eq. (\ref{PN}) is shown in Fig. \ref{fig:entgl}(b) for $N=11$. Fig. \ref{fig:entgl}(c) shows the behavior of $F_Q$ (continuous gray line) and $F_\mathrm{LSS}$ (black circles), which were numerically computed, as a function of $N$. The average percentage difference between $F_Q$ and $F_\mathrm{LSS}$ for the points shown in Fig. \ref{fig:entgl}(c) is $0.05\%$.

\subsection{Squeezed/coherent states at the interferometer inputs}

Let us now consider the seminal proposal of Caves of sending a squeezed vacuum state $\ket{\xi}_{a'}$ and a coherent state $\ket{\alpha}_{b'}$ at the inputs of the interferometer of Fig. 1(b) to improve the phase sensing precision \cite{caves81}. This procedure is currently used to improve the sensitivity of gravitational wave detectors \cite{ligo11,aasi13,zhao20,Sequino21}. In the Fock basis, these input states can be written as
\begin{equation}\label{sq-coh-antes}
	\ket{\xi}_{a'}=\sum_{m=0}^\infty S_{2m}\ket{2m}_{a'},\;\;\ket{\alpha}_{b'}=\sum_{n=0}^\infty C_{n}\ket{n}_{b'},
\end{equation}
where $S_{2m}=\sqrt{(2m)!}(-e^{i\theta_s}\tanh{r})^m/(2^mm!\sqrt{\cosh{r}})$ and $C_n=e^{-|\alpha|^2/2}{(\alpha e^{i\theta_c})^{n}}/{\sqrt{n!}}$. Note that for optimal conditions we must have $\theta_s-2\theta_c=0$. With the action of the transformations described in Eq. (\ref{BS}) due to the beam splitter BS$_1$ of Fig. 1(b), the quantum state $\ket{\Psi}$ of Eq. (\ref{psi0}) has coefficients
\begin{equation}\label{sq-coh}
\begin{split}
    &A_{N,k}=\sum_{m=0}^{N/2}\frac{i^{N-2m}C_{N-2m}S_{2m}}{\sqrt{2^{N}}}\sqrt{\binom{N}{2m}}\times\\&\times\sum_{q=0}^{N-2m}i^{-2q+k}\binom{N-2m}{q}\binom{2m}{k-q}\bigg/\sqrt{\binom{N}{k}},
\end{split}
\end{equation}
where for this state, to centralize the peak of the phase distribution in $\phi=0$, we considered that the phase $\theta_a$ after the BS$_1$ is $\theta_a=-\pi/2$. Note that in this case we do not have a definite total number of photons, as with the state $\ket{\Psi_N}$  of Eq. (\ref{psin}) that we are using to compute $F_Q$ and $F_\mathrm{LSS}$. So, to make comparisons we will project the state on a total number of photons equal to the average number of photons of the original state.

The maximum precision of this protocol occurs when the average number of photons of the squeezed state is equal to the average number of photons of the coherent state at the interferometer inputs, with $\sinh^2(r)=|\alpha|^2=\bar{N}/2$, $\bar{N}$ being the average number of photons of the total state \cite{pezze08}. Considering the quantum state at this configuration and projecting this state in a total number of photons $\bar{N}$, Fig. 7(a) shows the squared modulus of the coefficients $A_{N,k}$ from Eq. (\ref{sq-coh}) for $\bar{N}=10$. The relative phase distribution of Eq. (\ref{PN}) is shown in Fig. 7(b), again for $\bar{N}=10$. Fig. 7(c) shows the behavior of $F_Q$ (continuous gray line) and $F_\mathrm{LSS}$ (black circles), which were numerically computed, as a function of $\bar{N}$, always in a configuration with $\sinh^2(r)=|\alpha|^2=\bar{N}/2$. The traced red line shows the function $N^2$, showing that this scheme is really optimum, achieving the Heisenberg limit. The average percentage difference between $F_Q$ and $F_\mathrm{LSS}$ for the points shown in Fig. 7(c) is  0.05\%.

\begin{figure}
  \centering
    \includegraphics[width=8.5cm]{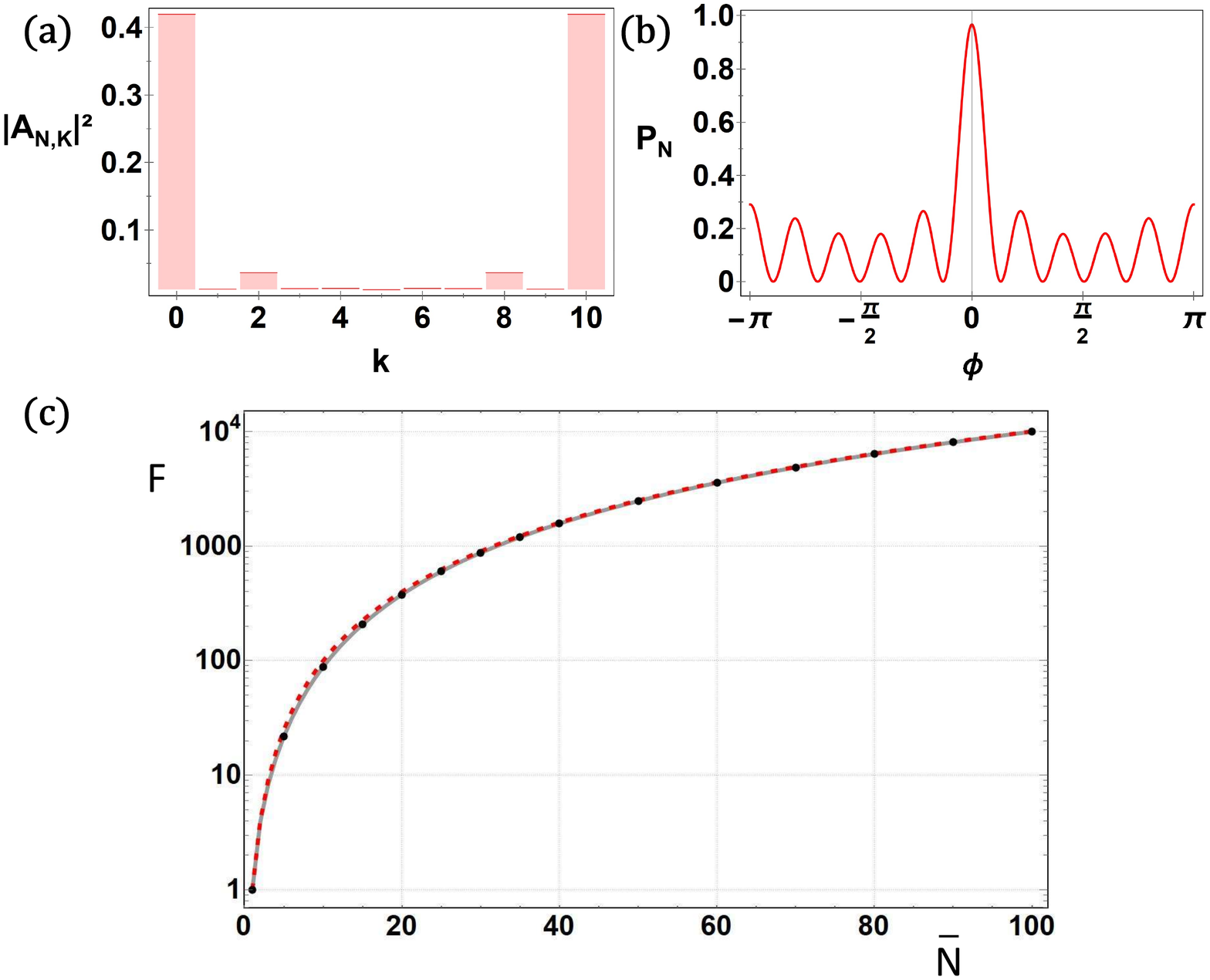}
  \caption{Analysis of a squeezed state and a coherent state from Eq. (\ref{sq-coh-antes}) sent at the inputs of the interferometer of Fig. 1(b), in the optimal configuration with $\sinh^2(r)=|\alpha|^2=\bar{N}/2$, with the state in modes $a$ and $b$ projected in a total number of photons $\bar{N}$. The resultant state is given by Eq. (\ref{psin}) with the coefficients of Eq. (\ref{sq-coh}). (a) Squared modulus of the coefficients from Eq. (\ref{sq-coh}) in the cited configuration with $\bar{N}=10$. (b) Relative phase distribution from Eq. (\ref{PN}) for $\bar{N}=10$. (c) $F_Q$ from Eq. (\ref{Fisher}) (continuous gray line) and $F_\mathrm{LSS}$ from Eq. (\ref{Fisher-LSS}) (black circles), both numerically computed, as a function of $N$. The traced red line shows the function $N^2$.}
\end{figure}

There is another regime where this configuration with a squeezed state and a coherent state at the input ports of the interferometer of Fig. 1(b) is very useful. If we have $\sinh^2(r)\approx\sqrt{\bar{N}}/2$ with $\bar{N}\gg1$, the information about the phase $\theta$ introduced in the interferometer can be extracted solely from the difference in the photon numbers detected by detectors D$_1$ and D$_2$ \cite{dobr15,polino20}. A disadvantage is that the minimum uncertainty in the estimated phase scales with $1/\bar{N}^{3/4}$ (the quantum Fisher information is proportional to $N^{3/2}$), not with $1/\bar{N}$ as in the optimum case. Again, we project the considered state in a total number of photons equal to $\bar{N}$ to obtain a state like the one of Eq. (\ref{psin}) with the coefficients $A_{N,k}$ given by Eq. (\ref{sq-coh}) in this regime with $\sinh^2(r)=\sqrt{\bar{N}}/2$. Fig. 8(a) shows the squared modulus of the coefficients $A_{N,k}$ from Eq. (\ref{sq-coh}) for $\bar{N}=10$. The relative phase distribution of Eq. (\ref{PN}) is shown in Fig. 8(b), again for $\bar{N}=10$. Fig. 8(c) shows the behavior of $F_Q$ (continuous gray line) and $F_\mathrm{LSS}$ (black circles), which were numerically computed, as a function of $\bar{N}$. The traced red line shows the function $1.45\bar{N}^{3/2}$. The average percentage difference between $F_Q$ and $F_\mathrm{LSS}$ for the points shown in Fig. 8(c) is $0.13\%$.

\begin{figure}
  \centering
    \includegraphics[width=8.5cm]{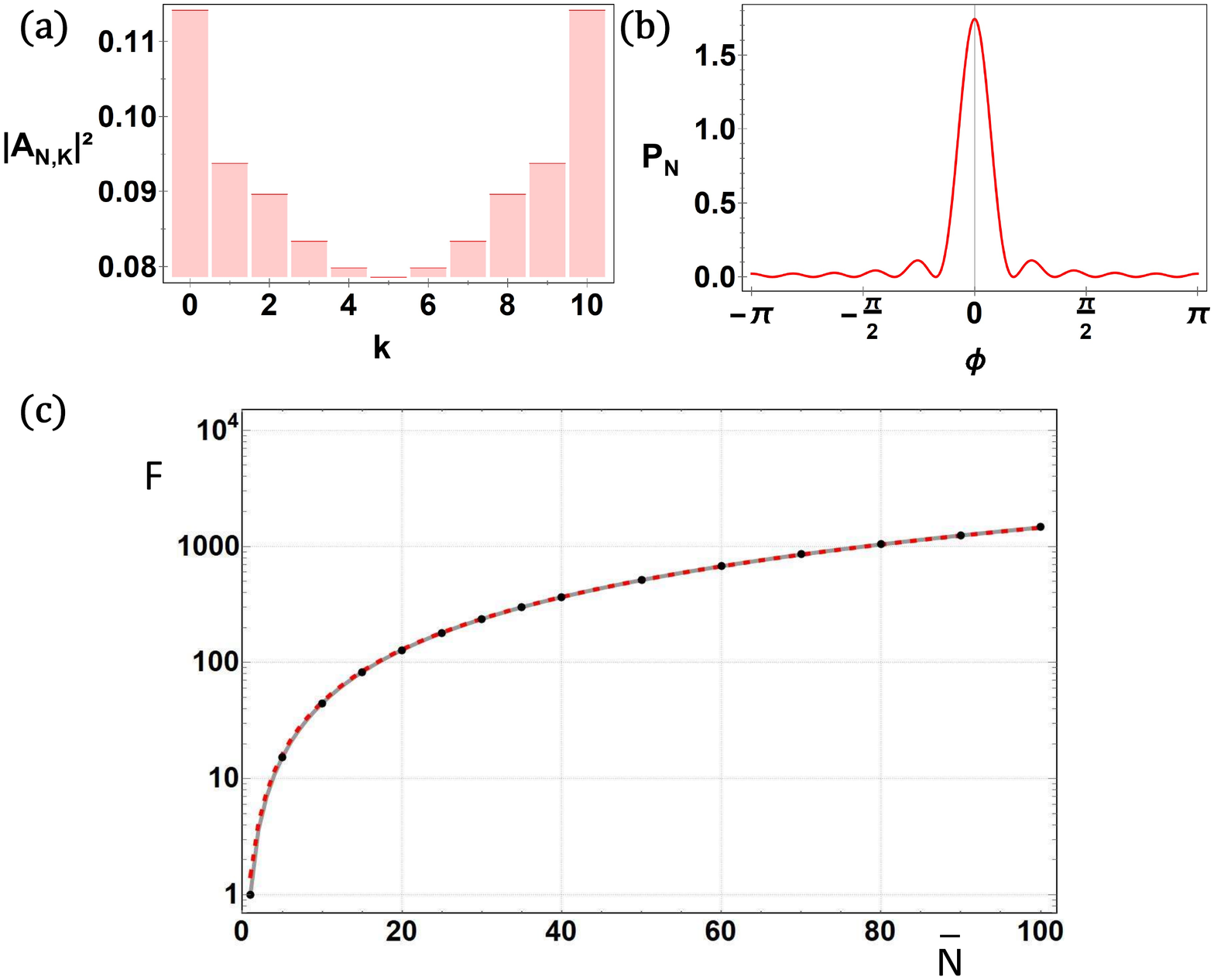}
  \caption{Analysis of a squeezed state and a coherent state from Eq. (\ref{sq-coh-antes}) sent at the inputs of the interferometer of Fig. 1(b), in a configuration with  $\sinh^2(r)=\sqrt{\bar{N}}/2$, with the state in modes $a$ and $b$ projected in a total number of photons $\bar{N}$. The resultant state is given by Eq. (\ref{psin}) with the coefficients of Eq. (\ref{sq-coh}). (a) Squared modulus of the coefficients from Eq. (\ref{sq-coh}) in the cited configuration with $\bar{N}=10$. (b) Relative phase distribution from Eq. (\ref{PN}) for $\bar{N}=10$. (c) $F_Q$ from Eq. (\ref{Fisher}) (continuous gray line) and $F_\mathrm{LSS}$ from Eq. (\ref{Fisher-LSS}) (black circles), both numerically computed, as a function of $N$. The traced red line shows the function $1.45\bar{N}^{3/2}$.}
\end{figure}

\section{Discussion and Conclusion}

To conclude, we have shown how the LSS relative phase distribution of many pure two-mode optical quantum states useful in quantum metrology is associated to the minimum uncertainty in the estimation of a small optical phase $\theta$ added to one of these modes. We have shown that the average difference between the Fisher information obtained from the LSS relative phase distribution and the quantum Fisher information for the tested states is less than 0.1\%, a difference compatible with the numerical precision of the calculations. This fact raises the question about if there is a formal equivalence between these quantities, and we are currently investigating this issue. 

We can see that, according to Eqs. (\ref{CRBmin}) and (\ref{Fisher}), a minimization of $\Delta\theta_\mathrm{min}$ is achieved by a maximization of $F_Q$ and, consequently, of $\Delta n_a$, which represents the standard deviation in the number of photons in mode $a$ in Fig. 1. To maximize $\Delta n_a$, the coefficients $A_{N,k}$ in Eq. (\ref{psin}) must have large amplitudes in the extremes $k=0$ and $k=N$. This is true for all quantum states useful in quantum metrology that we studied, notably for the NOON state, as seen in Fig. 3(a), and for the optimum state obtained with the incidence of a squeezed state and a coherent state at the interferometer inputs, as seen in Fig. 7(a), which are the ones that achieve the  Heisenberg limit $\Delta\theta_\mathrm{min}=1/(N\sqrt{p})$. The large components of the state with $k=0$ and $k=N$ leads to a high contribution of an oscillation term $\cos(N\phi)$ in the relative phase distribution, as can be seen in Figs. 3(b) and 7(b), such that the addition of a small phase difference appreciably changes the distribution, resulting in a large value for $F_\mathrm{LSS}$ in Eq. (\ref{Fisher-LSS}). These behaviors do not occur for a Fock state at one interferometer input, as evidenced by Figs. 2(a) and 2(b), and can be considered the responsible for the improvement of the metrology protocols with the use of quantum light states. Our results  illustrate the fact that, for a variation $\delta\theta$ in the quantity $\theta$ to be detectable, it must appreciably change the relative phase distribution of the initial quantum state.

In real practical situations in quantum metrology, where the initial state preparation is not perfect and the system is subjected to decoherence processes, the estimation of the protocol maximum precision based on the quantum Fisher information can be rather complicated \cite{escher11}. Perhaps the extension of the results presented here to mixed states of the two-mode quantum field may be useful for giving estimations for the protocol precision based on the LSS relative phase distribution of the optical quantum state.  This would result in an important practical application of the ideas introduced here.

The width of the LSS relative phase distribution of quantum fields was recently used to present a general fundamental explanation for the noise reduction in experiments that produce and characterize squeezed states of light \cite{calixto20}. If one takes into account the fact that the laser field is not a coherent state, but an incoherent combination of coherent states with random phases \cite{walls}, which is equivalent to an incoherent combination of Fock states \cite{molmer97,saldanha14}, the conclusion is that it is not possible to produce a single-mode squeezed state in the usual setups \cite{calixto20}. However, the width of the LSS relative phase distribution between the signal field and the field used as a local oscillator reduces with the increase of the squeezing parameter in the expected way \cite{calixto20}. This fact can be considered to be the fundamental reason for the noise reduction in the experiments. So, the analysis of the LSS relative phase distribution between two quantum optical modes may clarify other important aspects of quantum metrology, besides the ones treated here.

The authors acknowledge Leonardo Souza for useful discussions.  FFB acknowledges the N\'ucleo de Acessibilidade e Inclus\~ao da UFMG (NAI-UFMG) and the Pr\'o-Reitoria de Assuntos Estudantis da UFMG (PRAE-UFMG) for giving the necessary support for the realization of this research with his vision impairment. He also acknowledges all the monitors who helped him in this process: Raphaela de Oliveira, Tam\'iris Calixto, Jo\~ao Frossard, Ludmila Botelho, Thiago Carvalho, and Diego Ferreira. This work was supported by the Brazilian agencies CNPq, CAPES, and FAPEMIG.


%

\end{document}